# Social shaping of information infrastructure: on being specific about the technology

Eric Monteiro
Dept. of Informatics, Univ. of Trondheim, 7055 Dragvoll, Norway, Eric.Monteiro@ifi.unit.no

Ole Hanseth
Norwegian Computing Center, PB 114 Blindern, 0314 Oslo, Norway, Ole.Hanseth@nr.no

**Abstract**

We are in this paper discussing conceptualisations of the relationship between IT and organisational issues. To move beyond an "IT enables/ constrains" position, we argue that it is necessary to take the specifics of an information system (IS) more serious. A theoretical framework called actor-network theory from social studies of science and technology is presented as promising in this regard. With respect to new organisational forms, the class of ISs which need closer scrutiny is information infrastructures (INIs). They have characteristics which distinguish them from other ISs, namely the role and pattern of diffusion of standards. These standards are neither ready-made nor neutral: they inscribe organisational behaviour deeply within their "technical" details. Diffusion and adoption of standards depart from other kinds of ISs by requiring the coordination of the surrounding actors, institutional arrangements and work practices.

## 1 Introduction

The theme of this paper is theorising the relationship between technological and organisational issues, focusing on one such theory called actor-network theory (ANT) and contrasting it with Giddens' structuration theory. This analysis is then extended to the more restricted relationship between information infrastructure (INI) and new organisational forms.

We are witnessing a rapidly increasing number of theoretical, speculative or empirical accounts dealing with the background, contents and implications for a restructuring of private and public organisations. The sources of these accounts mirror the complex and many-faceted issues raised of economical (OECD 1992), social (Clegg 1990), political (Mulgan 1991) and technological (Malone and Rockart 1993; Malone, Yates and Benjamin 1991) nature. A comprehensive account of this is practically prohibitive; the only feasible strategy is to focus attention on a restricted set of issues. To explain how we intend to approach this, it is necessary to make certain assumptions explicit and indicate how we position our work relative to other accounts.

As our point of departure, we embrace the fairly widespread belief that IT is a, perhaps the, crucial factor as it simultaneously enables and amplifies the trends for restructuring of organisations (Applegate 1994; Orlikowski 1991). The problem, however, is that this belief does not carry us very far; it is close to becoming a cliche. To be instructive in an inquiry concerning current organisational transformations, one has to supplement it with a grasp of the interplay between IT and organisations in more detail. We need to know more about how IT shapes, enables and constrains organisational changes. Faced with such a situation, there are in principle two possible strategies: (i) a more proper understanding should emerge as the result of an appropriate theoretical framework (Orlikowski and Robey, p. 144) or (ii) through the accumulation of more empirical material (Smithson, Baskerville and Ngwenyama 1994, p. 10). Obviously, we neither want to argue against more suitable theoretical constructs nor empirical evidence per se. But we do suggest that what is lacking most is a satisfactory account of the interwoven relationship between IT and organisational transformations. More specifically, we argue that we need to learn more about how this interplay works, not only that it exists. This implies that it is vital to be more concrete with respect to the specifics of the technology. As an information system (IS) consists of a large number of modules and inter-connections, it may be approached with a varying degree of granularity. We cannot indiscriminatingly refer to it as IS, IT or computer systems. Kling (1991, p. 356) characterises this lack of precision as a "convenient fiction" which "deletes nuances of technical differences". It is accordingly less than prudent to discuss IS at the granularity of an artefact (Pfaffenberger 1988), the programming language (Orlikowski 1992), the overall architecture (Applegate 1994) or a media for communication (Feldman 1987). To advance our understanding of the interplay it would be quite instructive to be as concrete about which aspects, modules or functions of an IS enable or constrain which organisational changes — without collapsing this into a deterministic account (Monteiro, Hanseth and Hatling 1994).

The remainder of the paper is organised as follows. In section 2 we elaborate the argument sketched above, namely that scholarly studies need to be more specific about the various components of an IS. We critically discuss contributions to the conceptualisations of IT and organisations based on Giddens' structuration theory and argue that these studies are lacking in precision regarding the specifics of the IS. An alternative, theoretical framework borrowed from the field of social studies of science and technology called actor-network theory (ANT) is presented. ANT is on a fairly general and theoretical ground argued to offer a more promising way to account for the specifics of an IS. From this theoretical discussion concerning the appropriate way to conceptualise IT and organisations, we in section 3 turn to the issue of new organisational forms. The line of thought developed above here take on the following shape. Even brief reviews of arguments and evidence for new organisational forms motivate the need for a firmer understanding of the interplay with IT. The need to be specific and concrete regarding IT, in this context, amounts to describe how minute, technical design decisions embodied in the standards which constitute the information infrastructure are interwoven with organisational issues. The relevant class of IT to study is what during the last couple of years is called information infrastructure. Its technical back-bone is the standards. We accordingly study the social construction of standards. Our approach is based on two assumptions: (i) these standards are not ready-made, they are currently being shaped through complex, social processes, and (ii) these standards, contrary to most accounts, are nothing but neutral: buried deep in "technical" details they inscribe anticipations of individual, organisational and inter-organisational behaviour. Extending the more general argument of section 2, we suggest that ANT is quite useful in accounting for how standards acquire stability, how they become increasingly "irreversible". Section 4 presents empirical illustrations from two cases of how ANT, in a quite concrete manner, may be used to describe important and neglected aspects of INI. In section 5 we make a few comments on important aspects of the interplay between technical and non-technical elements of INIs which ANT cannot properly account for. In section 6 contains concluding remarks.

## 2 Conceptualisations of the relationship between IT and organisations

The problem of how to conceptualise and account for the relationship between, on the one hand, IT development and use and, on the other hand, organisational changes is complex — to say the least. A principal reason for the difficulty is due to the contingent, interwoven and dynamic nature of the relationship. There exists a truly overwhelming body of literature devoted to this problem. We will discuss a selection of contributions which are fairly widely cited and we which consider important. (Consult, for instance, (Coombs, Knights and Willmott 1992; Kling 1991; Orlikowski and Robey 1991; Walsham 1993) for a broader discussion.) Our purpose is to motivate a need to incorporate into such accounts a more thorough description and understanding of the minute, grey and seemingly technical properties of the technology and how these are translated into non-technical ones.

### 2.1 The duality of IT

The selection of contributions we consider all acknowledge the need to incorporate, in some way or another, that subjects interpret, appropriate and establish a social construction of reality (Galliers 1992; Kling 1991; Orlikowski 1991; Orlikowski and Robey 1991; Smithson, Baskerville and Ngwenyama 1994; Walsham 1993). This alone enables us to avoid simple-minded, deterministic accounts. The potential problem with a subjectivist stance is how to avoid the possibility that, say, an IS could be interpreted and appropriated completely freely, that one interpretation would be as reasonable as any other. This obviously neglects the constraining effects the IS have on the social process of interpretation (Akrich 1992; Bijker 1993; Orlikowski and Robey 1991). A particularly skilful and appealing elaboration of this insight is the work done by Orlikowski, Walsham and others building on Giddens' structuration theory (Orlikowski 1991, 1992; Orlikowski and Robey 1991; Walsham 1993).

Despite the fact that these accounts, in our view, are among the most convincing conceptualisations, they have certain weaknesses. These weaknesses have implications for the way we in later sections approach the question of the relationship between information infrastructure and new organisational forms. Our principal objection to conceptualisations like (Orlikowski 1991, 1992; Orlikowski and Robey 1991; Walsham 1993) is that they are not fine-grained enough with respect to the technology to form an appropriate basis for understanding or to really inform design. Before substantiating this claim, it should be noted that the studies do underline an important point, namely that "information technology has both restricting and enabling implications" (Orlikowski and Robey 1991, p. 154). We acknowledge this, but are convinced that it is necessary to push further: to describe in some detail how and where IT restricts and enables action. At the same time, we prepare the ground for the alternative framework of ANT describing the social construction of technology. To this end, we briefly sketch the position using structuration theory.

The aim of structuration theory is to account for the interplay between human action and social structures. The notion of "structure" is to be conceived of as an abstract notion; it need not have a material basis. The two key elements of structuration theory according to Walsham (1993, p. 68) are: (i) the manner in which the two levels of actions and structure are captured through the duality of structure, and (ii) the identification of modalities as the vehicle which link the two levels. One speaks of the duality of structure because structure constrains actions but, at the same time, human action serve to establish structure. This mutual interplay is mediated through a linking device called modalities. As modalities are what link action and structure, and their relationship is mutual, it follows that these modalities operate both ways. There are three modalities: interpretative scheme, facility and norm. An interpretative scheme deals with how agents understand and how this understanding is exhibited. It denotes the shared stock of knowledge which humans draw upon when interpreting situations; it enables shared meaning and hence communication. It may also be the reason why communication processes are inhibited. In applying this to IT, Orlikowski and Robey (1991, p. 155) notes that "software technology conditions certain social practices, and through its use the meanings embodied in the technology are themselves reinforced". Facility refers to the mobilisation of resources of domination, that is, it comprises the media through which power is exercised. IT, more specifically, "constitutes a system of domination" (ibid., p. 155). Norms guide action through mobilisation of sanctions. As a result, they define the legitimacy of interaction. They are created through continuous use of sanctions. The way this works for IT is that IT "codifies" and "conveys" norms (ibid., pp. 155 - 156).

Given this admittedly brief outline of the use of structuration theory for grasping IT, we will proceed by documenting in some detail how these accounts fail to pay due attention to the specifics of IT. Orlikowski and Robey (1991, p. 160) point out how "tools, languages, and methodologies" constrain the design process. The question is whether this lumps too much together, whether this is a satisfactory level of precision with regards to the specifics of IT. There are, after all, quite a number of empirical studies which document how, say, a methodology fails to constrain design practice to any extent; it is almost never followed (Curtis, Krasner and Iscoe 1988). Referring to Orlikowski (1991), Walsham (1993, p. 67) notes that "the ways in which action and structure were linked are only briefly outlined". This is hardly an overstatement as (Orlikowski 1991), as opposed to what one might expect from examining "in detail the world of systems development" (ibid., p. 10), maintains that the CASE tool — which is never described despite the fact that such tools exhibit a substantial degree of diversity (Vessey, Jarvenpaa and Tractinsky 1992) — was the "most visible manifestation" of a strategy to "streamline" the process (Orlikowski 1991, p. 14). (Orlikowski 1992) suffers from exactly the same problem: organisational issues are discussed based on the introduction of Lotus Notes. We are never explained in any detail, beyond referring to it as "the technology" or "Notes", the functions of the applications. This is particularly upsetting considering that Lotus Notes is a versatile, flexible application level programming language.

In instructive, in-depth case studies, Walsham (1993) does indeed follow up his criticism cited above by describing in more detail than Orlikowski (1991, 1992) how the modalities operate. But this increased level of precision does not apply to the specifics of the technology. The typical level of granularity is to discuss the issue of IBM vs. non-IBM hardware (ibid., pp. 92-94), centralised vs. decentralised systems architecture (ibid., p. 105) or top-down, hierarchical control vs. user-control (ibid., p. 136 - 138).

Not distinguishing more closely between different parts and variants of the elements of the IS is an instance of the aforementioned "convenient fiction" (Kling 1991, p. 356). An unintended consequence of not being fine-grained enough is removing social responsibility from the designers (ibid., p. 343). It removes social responsibility in the sense that a given designer in a given organisation obliged to use, say, a CASE tool, may hold that it is irrelevant how she uses the tool, it is still a tool embodying a certain rationale beyond her control.

## 2.2 Actor-network theory

What is required, then, is a more detailed and fine-grained analysis of the many mechanisms, some technical and some not, which are employed in shaping social action. We are not claiming that structuration theory cannot deliver this (cf. Walsham 1993, p. 67). But we are suggesting that most studies conducted so far (Korpela 1994; Orlikowski 1991, 1992; Orlikowski and Robey 1991; Walsham 1993) are lacking in describing, with a satisfactory level of precision, how specific elements and functions of an IS relate to organisational issues. We also suggest that the framework provided by actor-network theory (ANT) is more promising in this regard. We proceed to give an outline of the basics of ANT based on (Akrich 1992; Akrich and Latour 1992; Callon 1991; Latour 1987) before discussing what distinguishes it from the position outlined above.

Key concepts, for the present purposes, are: actor-network, translation, alignment, inscription and irreversibility. We explain them in order. ANT, not unlike structuration theory, recognises that establishing and changing a social order relies on a tight interplay between social and technical means; one speaks of society as a socio-technical web. The difference lies in how they describe this web. According to ANT, humans and non-humans are linked together into actor-networks. Further, ANT assumes that (a section of) society is inhibited by actors pursuing interests. An actor's interest can be translated into technical or social arrangements, for instance an IS or organisational routines. A basic question it sets out to explain is how a diverse group of actors reach agreement at all, that is, how a social order establishes a certain degree of stability or exhibits structural properties. According to ANT, stability is the end-result of the social process of aligning an initially diverse collection of interests to "one"; acceptance, "truth" or stability is the result of reaching a certain degree of alignment of interests (Callon 1991). The solution reached is constituted by an aligned actor-network. To achieve this, it is vital, as actors' interests from the outset are non-aligned, that one is successful in translating, that is, re-presenting or appropriating the interests of others to one's own (Latour 1987). The strength of ANT is very much related to exactly this point: ANT provides a language for describing how this translation takes place on a quite specific and concrete level. This takes us to the notion of an inscription (Akrich 1992; Akrich and Latour 1992).

An inscription is the result of the translation of one's interest into material form (Callon 1991, p. 143). In general, any component of the heterogeneous network of skills, practices, artefacts, institutional arrangements, texts and contracts establishing a social order may be the material for inscriptions. There are four especially interesting aspects of the notion of inscriptions: (i) what is inscribed, that is, which anticipations of use are envisioned, (ii) who inscribes them, (iii) how are they inscribed, that is, what is the material for the inscriptions and (iv) how powerful are the inscriptions, that is, how much effort does it take to oppose an inscription.

The inherent difficulty in changing an actor-network, that is, removing an inscription, can nicely be captured by Callon's concept of the (possible) irreversibility of an aligned network (Callon 1991, 1992, 1994). This concept describes how translations between actor-networks are made durable, how they can resist assaults from competing translations. Callon (1991, p. 159) states that the degree of irreversibility depends on (i) the extent to which it is subsequently impossible to go back to a point where that translation was only one amongst others; and (ii) the extent to which it shapes and determines subsequent translations.

### 2.3 Actor networks meet structuration theory

Having given an outline of ANT, let us turn to see what is achieved vis-à-vis structuration theory. The principal improvement, as we see it, is the ability ANT provides to be more specific and concrete with respects to the functions of an IS. It is not the case, in our view, that ANT in every respect is an improvement over structuration theory. We only argue that it applies to the issue of being specific about the technology. For instance, we consider the important issue of the structuring abilities of institutions to be better framed within structuration theory than within ANT. Let us explain why we consider it so. We first compare the two theories on a general level, partly relying on pedagogic examples. Then we attempt to reinterpret (Orlikowski 1991) in terms of ANT.

Inscriptions are given a concrete content because they represent interests inscribed into a material. The flexibility of inscriptions vary, that is, some structure the pattern of use strongly, others quite weakly. The power of inscriptions, that is, whether they must be followed or can be avoided, depends on the irreversibility of the actor-network they are inscribed into. It is never possible to know before hand, but by studying the sequence of inscriptions we learn more about exactly how and which inscriptions were needed to achieve a given aim. To exemplify, consider what it takes to establish a specific work routine. One could, for instance, try to inscribe the required skills through training. Or, if this inscription was too weak, one could inscribe into a textual description of the routines in the form of manuals. Or, if this still is too weak, one could inscribe the work routines by supporting them by an IS.

Latour (1991) provides an illuminating illustration of this aspect of ANT. It is a constructed example intended for pedagogic purposes. Hotels, from the point of view of management, want to ensure that the guests leave their keys at the front desk when leaving. The way this may be accomplished, according to Latour, is to inscribe this desired pattern of behaviour into an actor-network. The question then becomes how to inscribe it and into what. This is impossible to know before hand, so management makes a sequence of trials to test the strength of the inscriptions. First, they tried to inscribe it into an artifact in form of a sign behind the counter requesting all guests to return the key when leaving. This inscription, however, was not strong enough. Then one tried having a manual door-keeper with the same result. Management then inscribed it into a key knob. What they did was to use a metal knob of some weight. By stepwise increasing the weight of the knob, the desired behaviour was finally achieved. Through a succession of translations, the hotels' interest were finally inscribed into a network strong enough to impose the desired behaviour on the guests.

ANT's systematic blurring of the distinction between the technical and the non-technical extends beyond the duality of Orlikowski and Robey (1991) and Walsham (1993). The whole idea is to treat situations as essentially equal regardless of the means; the objective is still the same. Within ANT, technology receives exactly the same (explanatory!) status as human actors; the distinction between human and non-human actors is systematically removed. ANT takes the fact that, in a number of situations, technical artefacts in practice play the same role as human actors very seriously: the glue which keeps a social order in place is a heterogeneous network of human and non-human actors. A theoretical framework which makes an a priori distinction between the two is less likely to manage to keep its focus on the aim of a social arrangement regardless of whether the means for achieving this are technical or non-technical. The consequence of this is that ANT supports an inquiry which traces the social process of negotiating, redefining and appropriating interests back and forth between an articulate explicit form and a form where they are inscribed within a technical artefact. With reference to the small example above, the inscriptions attempting to establish the work routine were inscribed in both technical and non-technical materials. They provide a collection of inscriptions — all aimed at achieving the same effect — with a varying power. In any given situation, one would stack the necessary number of inscriptions which together seem to do the job.

We believe that the empirical material presented by Orlikowski (1991) may, at least partially, be reinterpreted in light of ANT. Her primary example is the development and use of a CASE tool in an organisation she calls SCC. The control (and productivity) interests of management are inscribed into the tool. The inscriptions are so strong that the consultants do as intended down to a rather detailed level. The only exceptions reported are some senior consults saying that they in some rare instances do not do as the tool require.

What is missing, then, in comparison with ANT is to portray this as more a stepwise alignment than the kind of all-in-one character of (ibid.). In ANT terms, the management's control interests are inscribed into the CASE tool in forms of detailed inscriptions of the consultants behaviour. The inscriptions are very strong in the sense that there is hardly any room for interpretive flexibility. The CASE tool is the result of a long process where management's control and productivity interests have been translated into a larger heterogeneous actor-network encompassing career paths, work guidelines, methodologies and, finally, the CASE tool. Together these elements form an actor-network into which consultants' behaviour are inscribed. Just like Latour's example presented above, the inscriptions become stronger as they are inscribed into a larger network. This network is developed through successive steps where inscriptions are tested out and improved until the desired outcome is reached. It is only when, as a result of a long sequence of testing and superpositioning of inscriptions, that one ends up in situations like the one presented by Orlikowski (1991). If one succeeds in aligning the whole actor-network, the desired behaviour is established. Analytically, it follows from this that if any one (or a few) of the elements of such an actor-network is not aligned, then the behaviour will not be as presented by Orlikowski (ibid.). Empirically, we know that more often than not the result is different from that of Orlikowski's case (Curtis, Krasner and Iscoe 1988; Vessey, Jarvenpaa and Tractinsky 1992).

We end this section by merely pointing out another issue we find problematic with (Orlikowski 1992). She states that "[t]he greater the spatial and temporal distance between the construction of a technology and its application, the greater the likelihood that the technology will be interpreted and used with little flexibility. Where technology developers consult with or involve future users in the construction and trial stages of a technology, there is an increased likelihood that it will be interpreted and used more flexibly" (ibid., p. 421). We agree on the importance of user participation in design. According to ANT, however, the interpretive flexibility of a technology may increase as the distance between designers and users increases. Interpretive flexibility means unintended use, i.e. using the technology different for what is inscribed into it. When the designers are close to the users, the network into which the intended user behaviour is inscribed will be stronger and accordingly harder for the users not to follow this. An important aspect of ANT is its potential to account for how restricted interpretative flexibility across great distances can be obtained (Law 1986).

## 3  New organisational forms and information infrastructure

### 3.1  Some of the arguments — and the evidence

New organisational forms are assumed important in order to achieve enhanced productivity, competitiveness, flexibility, etc. New organisational forms are usually of a network type positioned between markets and hierarchies. The discussions about new organisational forms borrow from a number of sources. We briefly review some of these.

From economics, basically relying on transaction-cost considerations, there is a growing pressure to accommodate to the "information economy" (Ciborra 1992; OECD 1992). Transaction-cost considerations fail to do justice to the dynamically changing redivision of labour and functions which are two important aspects of new organisational forms (Ciborra 1992). Within business policy literature the arguments focus on issues of innovation processes as facilitated through strategic alliances and globalization which emerge pragmatically from concerns about maintaining competitiveness in a turbulent environment (Porter 1990; von Hippel 1988). In organisational theory, one emphasises the weaknesses of centralised, bureaucratic control in terms of responsiveness to new situations (Clegg 1990). Ciborra (1992) sees new organisational forms as rational, institutional arrangements to meet the increased need for organisational learning. Technological development within information and communication technology are identified by some scholars as the driving force for the restructuring of organisations (Malone, Yates and Benjamin 1991; Malone and Rockart 1993).

Even such a brief exposition of theoretical considerations should make it evident that the issue of new organisational forms is vast. When we turn to what exists of empirical evidence, the picture gets even more complicated. This is because the empirical material document a far less clear-cut picture as it contains numerous contradicting trends (Applegate 1994; Capello and Williams 1992; Orlikowski 1991; Whitaker 1992). In line with the argument above, we intend to approach matters in a highly selective manner. As in the more general case discussed in section 2, the principal challenge is to account convincingly for the interwoven relationship between new organisational forms and its IT based back-bone, that is, INI. We are hence led to consider in more detail the specifics of an INI.

### 3.2  Information infrastructure

INIs have a number of characteristic features which serve to distinguish it from other kinds of ISs. These differences are not only of contrived interest. They have strong repercussions on the way INIs are developed, spread and used. We point out and discuss a few of the characteristics which need a more proper understanding.

A principal difference, related to the fact that an INI is a systemic technology which regulates communicative behaviour, is the role and status of standards. For most technologies including the bulk of other ISs, standards evolve gradually as the technology matures. What makes INIs different is the absolute requirement that all involved parties have to adhere to a standard at any given moment. The INI simply ceases to exist if communication does not follow the standard (Besen and Saloner 1989).

Another characteristics of INI, related to that above, is the way it acquire its stability. Because any modifications of the standards need to be coordinated and organised to avoid collapsing the communicative behaviour, modifying a standard grows increasingly more difficult as the standard diffuses. In an economical vocabulary, this property of an INI produces "network externalities" (Antonelli 1993), that is, a situation where the value for the users increases with the diffusion of the technology creating lock-ins and self-reinforcing effects. Expanding the installed base of a standard gives rise to an accumulated "momentum" of the standard. This aspects is nicely captured by the notion in ANT of "irreversibility" (Callon 1991) thus providing an additional and different motivation for ANT than the ones already given.

## 4 Information infrastructure as actor-network

Intuitively, as ANT's primary objective is to describe technological systems and non-technical structures as single units, i.e. as socio-technical networks, it should be well suited for describing interrelations between network organisations and network technologies.

We will in this section describe what we think are important aspects of INIs and show how ANT can help us understand and deal with them. These aspects are:

1. Standards, being the essence of INIs, are not just neutral technological components. Rather they inscribe their use, i.e. interorganisational communication patterns as well as actions that has to take place locally in user organisations.
2. Technological and non-technological elements are linked, implying (among other things) that which organisations are involved in the standardisation/INI design process shapes the standard as well as the standard has implications for which organisations need to participate in the design process. The actor-networks of INIs and INI standardisation easily turn into unmanageable complex ones.
3. A standard is more that just a technical system, it is a socio-technical network.
4. As standards become realised INIs they easily turn into irreversible actor networks.

The examples we are using to illustrate these aspects are primarily chosen from the development of an INI for the Norwegian health care sector. The examples are exchange of drug prescriptions and laboratory orders and results. Our presentation is based on a study of historical material (minutes, reports, proposed standards) and interviews, and on experience obtained by one of the authors engaged by one of the involved software vendors for a period of two years. Developing Norwegian standards for the health sector takes place in close connection with the development of European ones organised by CEN.

### 4.1 Inscribing inter-organisational behaviour into standards

The definition of the Norwegian standard for drug prescription exchange was organised as a project where those considered relevant participated. The main task of the project was the definition of an EDI message representing a prescription. An important part of such a message is the definition of how the prescribed drugs are identified. Early in the project, the pharmacies expressed their wish to use the same drug identification numbers in the electronic drug prescription message as the ones they are using in their existing systems. The pharmacies wanted these identification numbers because they could then integrate the system receiving prescriptions with their existing applications for inventory control and electronic ordering from The Norwegian Drug Medication Depot (NMD).

None objected to the suggestion to include this number in the message. All agreed because this information did not seem to conflict with anyone's interest. The problem was only that the GPs made no use of it and hence had no access to it in their work. Neither the GPs nor their applications were aware of which identifiers corresponded to which drugs. The initially straightforward and technical issue of how to code and represent a prescribed drug identification number — a six digit number — in the message had thus been translated into another issue: how should one inscribe this into the message in such a way that the GPs will provide this number as part of the prescription without creating additional work for the GPs? If using the drug identifiers as intended turns out to be inconvenient for the GPs, they will probably not do it properly. Proper use of the identifiers may be supported by the GPs' information system. This support may be offered by installing the drug item list as a part of the GPs information system. An alternative solution would be the drug item list integrated with the Common Catalogue (in Norwegian: Felleskatalogen). These two alternative technical solutions represent two alternative translations of the pharmacies interests.

The pharmacies' drug item list is provided by the NMD, and they make it available to the pharmacies via their application supplier. The list contains information useful also for the GPs, for instance about price and synonymous drugs. The GPs' representative argued that this list had to be made available for the GPs' applications. Today the GPs use their own drug lists which are either typed in by themselves or installed with the application (as a supplier's service). In either case, the GPs themselves have to update their drug list.

As everyone agreed that the GPs need the drug item list, the focus has been on whether or not the GPs should pay for access to the drug item list, and if so at which price. The Pharmacies' Association has recognised the GPs' need for the list. They have no principal objections against offering relevant parts of it to the GPs. The NMD, however, who is responsible for the drug item list, has not yet decided what to do. NMD is not represented in the project and will probably not offer the list free of charge. The pharmacies are not willing to let the GPs have access to the drug item list as it is at the present, because the list also contains information which the pharmacies want to keep for themselves (for instance about profit margins on pharmaceutical products). The list thus has to be tailored to the needs of the GPs. The drug item list is updated every month, each time the version used by GPs has to be produced. The vendors of the GPs' medical record system need to be adapted to make use of the list. The lists can be distributed to GPs directly from NMD or through the vendors of the GPs' systems.

The GPs also have available a paper based catalogue, called the Common Catalogue, containing information about all registered drugs in Norway. It also contains other important and practical information about treatment of acute poisoning, drug interactions, a register of drug producers and a register of all pharmacies in Norway. Having access to this catalogue as a part of their information systems would be very useful when the GPs are specifying a drug prescription. This catalogue is provided by yet another organisation, an organisation not part of the drug prescription project either. The catalogue is printed once a year, but additions are printed during the year as new drugs are introduced or disappear. The Common Catalogue now exists electronically as well. Work is being done to integrate the catalogue with the GPs' drug list to ease the work for the GPs. This is seen as a possible solution to the problem above, that is, how to offer something to the GPs which makes it acceptable for them to register the proper prescribed drug identification number. This integration work requires the cooperation of GPs, suppliers of GP systems, the NMD and the organisation responsible for the Common Catalogue. All these parties have commercial interests in this area, and their motivation must be combined with GPs' actual need for integration of drug information. This work has just started, and in the mean time the drug prescription exchange project has to solve the problem with how to make the item prescription number available for the GPs.

This example illustrate how rather complex inter-organisational behaviour related to the updating and distribution of a list of identifiers are inscribed into the definition of one single element of a standardised messages. Defining identifiers of this kind is usually considered a mere "technical detail". This example illustrates that this is not always the case. The technical design represents translated interest, in particular the pharmacies'. The identifier list is linked to a number of organisations, constituting an actor-network. This network must be aligned, i.e. each of the organisations must do as prescribed if the drug prescription exchange system shall work. The definition of identifiers presupposes a certain organisational behaviour. But this inscription may be too weak. To really make the GPs behave as intended, the inscriptions of their behaviour were strengthened by offering additional services as part of the system.

### 4.2 The interdependencies between standard and standardisation process

The example above also illustrates the interdependencies between standard and standardisation process. Those involved tries to translate their interests into the standard just like the pharmacies did. And a specific solution proposal may affect others in way making it necessary to involve them in the process as happened with NMD.

An commonly applied criteria for evaluating technical solutions is that they should be as simple as possible. From an ANT perspective this means that not only the technical system but rather the actor-network it is a part of should be as simple as possible. The more complex an actor-network, the more complex it is to align it. We believe that the actor-networks representing INIs may often be very complex, in fact almost too complex to be aligned. This will be illustrated by the development of EDI systems for exchange of lab results from labs to GPs.

The development of an INI for exchange of lab information in Norway started when Dr. Fürst's Medisinske Laboratorium, a privately owned chemical-clinical laboratory, developed a system enabling their customers (GPs) to receive the test results electronically. Fürst had for a long time been a competitive, service oriented lab and advanced in use of technology. They saw a strategic opportunity and developed a simple system. They paid the patient record vendors to adapt their products to it, and offered it free of charge to GPs (even paid their expenses to modems!). Receiving electronic reports from clinical-chemical labs makes the work of GPs significantly easier as each GP receives approximately 20 reports a day, which take quite some time to register manually in their medical record system. Fürst was very successful as the INI helped them getting a large number of new customers.

The competitive advantage Fürst obtained by means of their INI created a problem for other labs. Most of them are hospital labs and their expenditures are paid for by the owners, that is, the county authorities. The county pays also when a GP orders a test at a private lab. The profit margins are significant so the counties soon saw the need to respond by providing similar services. Most of them did that. The systems supporting the services were to a large extent copies of Fürst's.

Those involved in the development of these smaller INIs recognised that they should be considered parts of a larger INI for the health sector. This demands shared standards. Working out and spreading a stable standard amounts to successfully aligning the heterogeneous network of involved actors, institutions and work practises into a "convergent" network (Callon 1991). The effort of aligning such a network grows dramatically as the complexity, that is, the number of elements, of the network grows. Fürst's solution was simple and tailor-made to their needs. In particular, it involved a relatively small number of elements in the network. A general, standardised solution, however, would be considerably more complex. Figure 1 indicates this by connecting the elements in addition to Fürst's solution with dashed lines. It follows from this that aligning the network for the standardised solution is radically more difficult that for Fürst's.

*Fig.1 The actor-network of lab information exchange*

The increased complexity of a standardised solution is due to the presence of a large variety of labs (for instance, clinical-chemical, micro-biology, pathology and allergy), the ambition of also enable transmission of information among other institutions within health care (for instance, hospitals, various governmental institutions and health insurance authorities) and because it was decided that the Norwegian standardisation work should follow European standardisation work. It has been decided, for instance, that lab results should be exchanged as EDIFACT messages. This implies that the specific messages should fulfil the requirements to messages set by the EDIFACT standardisation bodies, and that these bodies (at both national and European level) will be involved in the standardisation work as well.

### 4.3 Designing a socio-technical network

The definition of a standard is more than designing a technological system. It is the design of a socio-technical network. The standard is just one element, an element linked to a number of other elements, technical and non-technical, in this network. This can be illustrated by the proposed systems for exchange of lab orders. One crucial issue here is to ensure that the electronic order is linked to its proper physical specimen at the lab. Today the paper order is physically connected to the specimen container and sent together from the GP to the lab. A unique number is glued to both. Lists of pairs of adhesive labels containing unique numbers are pre-produced by the labs and given to GPs. These lists are produced in a way minimising the probability of mixing up the labels on the order and the specimen container. Finding an equally safe procedure when the order is transmitted electronically will not be easy, and will certainly include the design of specific technological as well as organisational arrangements. This amounts to recognising that design of a solution for lab orders invariably involves the design of a large collection of associated work routines as well as computer systems. The solution must include label producing machines (bar code printers), label reading machines, manual routines and new computer applications. The standardised message will reflect the working routines, they will be inscribed into the message. For instance, what kind information is necessary for carrying out the control routines depends on how these routines are defined. This information must, of course, be represented in the message.

### 4.4 The irreversibility of the installed base

INIs and their standards must just as any information system change to adopt to local needs, changed needs, and to improve according to use experience (Hanseth, Thoresen and Winner 1993, Monteiro, Hanseth and Hatling 1994). Callon (1991, p. 159) states that an actor-network may turn irreversible depending on (i) the extent to which it is subsequently impossible to go back to a point where that translation was only one amongst others; and (ii) the extent to which it shapes and determines subsequent translations.

We believe that INIs easily may become irreversible as they grow. This is so because everybody communicating has to use the same standard. A standard may be changed by developing a new version which is compatible with the old one in the sense that all implementations of the old version may communicate with installations of the new as if they were equal. In this case, individual users can switch to the new version independently. However, using this strategy, what kind of changes that can be introduced into the new version is significantly limited. Defining a new incompatible version removes this constrain. On the other hand, switching to the new version is more difficult. One alternative is that all users switch over to the new version at the same time, at a so-called "flag day." Coordinating such a change, however, becomes difficult as the number of installations grows. A second strategy for switching to a new incompatible version may be to develop and install gateways between the old and the new. The difficulties in doing this depends on the degree of incompatibility between the two versions. A third strategy is to build up a new separate INI based on the new version. The drawback with this alternative is that the usefulness of an INI depends on the number of users connected. Accordingly, all users would prefer the old INI with many users rather than new one with very few.

To our knowledge, the INI for the Norwegian health care sector has not yet reached a state where its (possible) irreversibility has appeared as a problem. However, considering the difficulties in developing the first versions of the various standards, we assume that defining new versions will be harder as a large number of installed standardised systems in itself put additional constraints on what can be a suitable solution. In the Internet world, work has been going on since 1990 to develop a new version of the IP protocol. A new version with only modest changes from the existing one is expected to be settled late this year (RFC 1995). This new version requires changes in 58 other Internet standards, and a vast number of implementations of these Internet standards must be modified.

A number of more substantial changes are necessary if Internet shall be able to support, for instance, multimedia information and mobile communication properly. Necessary changes for these purposes were intended to be a part of the ongoing revision. The modest changes obtained during a five period indicates that Internet is approaching a state of irreversibility (Monteiro, Hanseth and Hatling 1994).

## 5  Beyond actor-networks

ANT cannot account for all relevant aspect of INIs. For instance, it cannot properly deal with institutions, i.e. how they shape actions at the same time as the very same actions shape the institutions, i.e. what we see as the primary contribution of structuration theory. The institutional framework of standardisation and INI development is an important part of these processes.

Another important aspect of INIs which ANT neither can account for properly is their openness. An INI is open in the sense that whatever its scope is, several user organisations would prefer the INI extended to other areas which they see as tightly connected with those areas where they are using it. In case of lab communication, the GPs communicate with several different kinds of labs. The labs are not only communicating with GPs but with hospitals, other labs etc. GPs communicate also, in a similar way as with labs, with X-ray clinics, hospitals, etc. Each of these institutions communicate with others, and so on indefinitely. Defining a unified, compatible set of standards for all is impossible. What is needed is a conceptual framework for developing flexible standards and INIs for limited areas which can be linked together in one way or another. I terms of ANT, what is needed is concepts for dealing with not only one, but larger numbers of actor-networks and how they connect and interact. Leigh Star's concept of boundary objects is a step in this direction (Star and Griesemer 1989).

## 6  Conclusion

There is in 1995 quite a widespread agreement among scholars in the field that "IT enables/ constrains action". It seems to use that the time is ripe to push further and learn more about how this operate on a concrete level. We argue that a firmer grasp of the interplay between IT and organisational issues have to take the specifics of an IS quite seriously. This is, in particular, necessary for informing design and increasing social responsibility (Kling 1991, p. 343).

Even highly appraised accounts (we have only considered a few which build on structuration theory but believe that it holds also more generally) fail in this crucial regard. This should not be read as a claim that structuration theory, in principle, is not capable of supporting this kind of concrete and specific description. But we do argue that ANT quite immediately provides a language for doing this; its overall rationale is geared towards this.There are, however, other aspects which, to us, seem to be better captured within structuration theory than ANT (for instance, the role of institutions). Given the premise that we need to be more specific, it strikes us as a more promising strategy to employ ANT than to await an elaboration of the notion of modalities, in ways not entirely clear to us, before proceeding.

With regards to new organisational forms, this implies that we need a thorough understanding of the characteristics of the relevant class of ISs, namely INIs and their underlying standards. This amounts, we argue, for a study both of how organisational behaviour is inscribed into "technical" details of the standard and how adoption and diffusion of a standard involve making it irreversible by aligning the surrounding, heterogeneous network of institutional arrangements and work practices.